\documentstyle[preprint,aps,prl]{revtex}  
\begin{document}
\title{ Abelian bosonization approach to quantum impurity problems }
\author{Jinwu  Ye}
\address{
Physics Laboratories,       
Harvard  University, Cambridge, MA, 02138}
\date{\today}
\maketitle
\begin{abstract}
  Using  Abelian Bosonization, we develop a simple and powerful method
 to calculate the correlation functions of the two channel Kondo model
 and its variants. The method can also be used to identify {\em all} the
 possible boundary fixed points and their {\em maximum} symmetry,
 to calculate straightforwardly the finite size spectra, to demonstrate
 the physical picture at the boundary explicitly.  Comparisons with 
 Non-Abelian Bosonization method are made.  Some fixed points corresponding
 to 4 pieces of bulk fermions coupled to $ s=1/2 $ impurity are listed.
\end{abstract}
\narrowtext

 In general quantum impurity problems, a local quantum mechanical degree of freedom
 couples locally to some extended degree of freedoms which can be described
 by some critical theory in the continuum limit. Some noted examples are the
 multichannel Kondo effects \cite{blandin}, quantum dissipation problems \cite{schmid},
 an impurity in one-dimensional Luttinger liquid \cite{kane} and the
 catalysis of proton decay \cite{callan}.  
 For general quantum impurity model, Affleck and Ludwig (AL) \cite{affleck},
 using CFT, pointed out that the impurity degree of freedoms completely {\em disappear} from
 the description
 of the low temperature fixed point and leave behind  conformally 
 invariant {\em boundary conditions}. For the multichannel Kondo problems,
 AL calculated the finite size
 spectrum at the fixed point which are consistent with the results
 of Numerical Renormalization Group \cite{nrg,twoimp}.
 AL calculated not only all the thermodynamic quantities, but also all
 the correlation functions \cite{resistivity,corre}.

 For the special case of 4 pieces of bulk fermions such as the two channel Kondo effect (2CK),
 the one channel two impurity Kondo models \cite{twoimp} and the four flavor Callan-Rubakov effect \cite{callan},
 the non-interacting theory possesses $ SO(8) $ symmetry,
 Maldacena and Ludwig (ML) \cite{ludwig} showed that
 the boundary conditions at the fixed point turn out to be {\em linear} in the basis
 which separates charge, spin and flavor.  The {\em linear}
 boundary conditions can also be transformed into the {\em non trivial}
  boundary conditions in the original
 fermion basis by the triality transformation \cite{ludwig}.
  By using the {\em linear} boundary conditions, ML reduced the problem of calculating all the 
 correlation functions to  free field exercises.
 However, ML deduced the boundary conditions by using
 explicitly the results of AL ( for example, the single particle S matrices vanish),
 therefore their method explicitly depends on AL's results. 

  Emery and Kivelson (EK) \cite{emery}, using Abelian
  Bosonization, found an alternative solution to the 2CK. Although EK's method
  is applicable only when the number of channels takes some special values,
  it is simpler and more widely accessible
  than AL's method.  The operator contents of AL's and EK's approaches
  have been shown to be exactly the same by the author \cite{line}.
  However, so far, EK's method cannot be used to calculate
  the {\em single-electron} properties.

    In this paper, using Abelian Bosonization,  we develop a simple, systematic and powerful method
 to study the 2CK and its variants. The method can
 identify {\em all} the possible boundary fixed points, their corresponding
 boundary conditions and their {\em maximum} symmetry in a very straightforward way.
 It can also calculate the finite size spectrum and {\em any} correlation functions.
 We explicitly point out that at the fixed point, 
 the impurity degree of freedoms act as {\em Lagrangian multipliers},
 therefore impose boundary conditions on the conduction electrons, the impurity themselves
 turn into the corresponding scaling fields at the fixed point.
 The method has also
 been successfully applied to study the two channel spin-flavor Kondo model \cite{sf},
 the two channel flavor anisotropic and one channel compactified Kondo models \cite{flavor}
 and the problem of a non-magnetic impurity hopping between two sites
 in a metal \cite{hopping}. Although demonstrating the method explicitly only by solving the 2CK,
 we suggest that our method maybe applicable to other quantum impurity problems.
 Comparisons with AL's and ML's results are made. Finally,
 some fixed points corresponding to 4 pieces of bulk fermions coupled to
 $ s=1/2 $ impurity are listed.

 The Hamiltonian of the well-studied 2-channel Kondo model is:
 \begin{equation}
 H =  i v_{F} \int^{\infty}_{-\infty} dx 
   \psi^{\dagger}_{i \alpha }(x) \frac{d \psi_{i \alpha }(x)}{dx}
   + \sum_{a=x,y,z} \lambda^{a} J^{a}(0)  S^{a} 
   +  h ( \int dx J^{z}_{s}(x) + S^{z} )
\label{kondob}
 \end{equation}
   where $ J^{a}(x) =\frac{1}{2} \psi^{\dagger}_{i \alpha }(x)
   \sigma^{a}_{\alpha \beta} \psi_{i \beta }(x) $ 
   is the  spin  currents of the conduction electrons.

  In this paper, for simplicity, we take $ \lambda^{x}=\lambda^{y}=\lambda$,
  the symmetry in the spin sector is  $ U(1) \times Z_{2} \sim O(2) $.
 Abelian-bosonizing the four bulk Dirac fermions separately:
\begin{equation}
 \psi_{i \alpha }(x )= \frac{P_{i \alpha}}{\sqrt{ 2 \pi a }}
  e ^{- i \Phi_{i \alpha}(x) }
\label{first}
\end{equation}
    Where  $ \Phi_{i \alpha} (x) $ are the real chiral bosons,
    the cocyle factors have been chosen as: $ P_{1 \uparrow}= P_{1 \downarrow}
  = e^{i \pi N_{1 \uparrow} }, P_{2 \uparrow}= P_{2 \downarrow}
  = e^{i \pi ( N_{1 \uparrow} + N_{1 \downarrow} + N_{2 \uparrow}) } $.

%satisfying the commutation relations
%\begin{equation}
%  [ \Phi_{i \alpha} (x), \Phi_{j \beta} (y) ]
%  =   \delta_{i j} \delta_{\alpha \beta} i \pi sgn( x-y )
%\end{equation}

   Following the three standard steps of the EK's solution \cite{emery},
  (1) Introduce charge, spin, flavor, spin-flavor bosons
  (2) Make the canonical transformation $ U= \exp [ i S^{z} \Phi_{s}(0)] $
  (3) Make the following refermionization

\begin{eqnarray}
S^{x} &= & \frac{ \widehat{a}}{\sqrt{2}} e^{i \pi N_{sf}},~~~
S^{y}= \frac{ \widehat{b}}{\sqrt{2}} e^{i \pi N_{sf}},~~~
S^{z}= -i \widehat{a} \widehat{b}        \nonumber \\
 \psi_{sf} & = & \frac{1}{\sqrt{2}}( a_{sf} - i b_{sf} ) =
  \frac{1}{\sqrt{ 2 \pi a}} e^{i \pi N_{sf}} e^{-i \Phi_{sf} }     \nonumber   \\
 \psi_{s,i} & = & \frac{1}{\sqrt{2}}(  a_{s,i} - i  b_{s,i} )=
 \frac{1}{\sqrt{ 2 \pi a}} e^{i \pi d^{\dagger}d} e^{i \pi  N_{sf}} e^{-i \Phi_{s} }   
\end{eqnarray}

%\begin{eqnarray}
%\Phi_{c} & = & \frac{1}{2} ( \Phi_{1 \uparrow }+ \Phi_{1 \downarrow }+
% \Phi_{2 \uparrow }+ \Phi_{2 \downarrow } )   \nonumber \\
%\Phi_{s} & = & \frac{1}{2} ( \Phi_{1 \uparrow }- \Phi_{1 \downarrow }+
% \Phi_{2 \uparrow }- \Phi_{2 \downarrow } )   \nonumber \\
%\Phi_{f} & = & \frac{1}{2} ( \Phi_{1 \uparrow }+ \Phi_{1 \downarrow }-
% \Phi_{2 \uparrow }- \Phi_{2 \downarrow } )   \nonumber \\
%\Phi_{sf}& = & \frac{1}{2} ( \Phi_{1 \uparrow }- \Phi_{1 \downarrow }-
% \Phi_{2 \uparrow }+ \Phi_{2 \downarrow } )
%\label{second}  
%\end{eqnarray}

  The transformed Hamiltonian $ H^{\prime}= U H U^{-1} =
    H^{\prime}_{sf} + H^{\prime}_{s} + \delta H^{\prime} $ can be written in terms of
    the Majorana fermions as \cite{atten}: 
\begin{eqnarray}
 H_{sf} &= & \frac{ i v_{F} }{2} \int dx (a_{sf}(x) \frac{ \partial a_{sf}(x)}
 {\partial x} + b_{sf}(x) \frac{ \partial b_{sf}(x)} {\partial x} )
   -i \frac{ \lambda }{\sqrt{ 2 \pi a}} \widehat{a} b_{sf}(0)
                                                 \nonumber \\
 H_{s}& = & \frac{ i v_{F} }{2} \int dx (a_{s}(x) \frac{ \partial a_{s}(x)}
 {\partial x} + b_{s}(x) \frac{ \partial b_{s}(x)} {\partial x} )
        -i h \int dx a_{s}(x) b_{s}(x)   \nonumber  \\
  \delta H &= & -\lambda_{z}^{\prime} \widehat{a} \widehat{ b} a_{s}(0) b_{s}(0)
\label{start}
\end{eqnarray}
  where $ \lambda_{z}^{\prime} = \lambda^{z} - 2 \pi v_{F} $. 

  From the above equation, it is evident that along the solvable line (EK line)
  $ \lambda^{\prime}_{z}=0 $, half of the impurity spin $\widehat{b}$ and $ \psi_{s}(x) $
  completely decouple \cite{emery}.
   However, the canonical transformation $ U $ is a boundary condition
   changing operator \cite{boundary},
  the transformed field $\psi^{\prime}_{s}(x) $ is related to
 the original field $ \psi_{s}(x) $ by
\begin{equation}
\psi^{\prime}_{s}(x) = U^{-1} \psi_{s,i}(x) U= e^{i \pi d^{\dagger} d} e^{i \pi S^{z} sgn x }
\psi_{s}(x) = -i sgn x \psi_{s}(x) 
\end{equation}

  It is important to observe that the impurity spin {\em completely disappear} from
 the above equation.

  The boundary condition $ \psi^{\prime}_{s,L}(0)=\psi^{\prime}_{s,R}(0) $ in $ H^{\prime} $ leads to
\begin{equation}
  a_{s,L}(0)=-a_{s,R}(0), ~~ b_{s,L}(0)=-b_{s,R}(0)
\label{bound1}
\end{equation}

   In order to identify the fixed point along the solvable line, 
  it is convenient to write $ H_{sf} $ in the action form
\begin{equation}
 S =  S_{0} + \frac{\gamma}{2} \int d \tau \widehat{a}(\tau)
       \frac{\partial \widehat{a}(\tau)}{\partial \tau}
      -i \frac{ \lambda }{\sqrt{ 2 \pi a}}
        \int d \tau \widehat{a}(\tau) b_{sf}(0,\tau)
\label{action}
\end{equation}

  The simple power countings of the action $ S $ leads to the following R. G. flow
  equations \cite{fisher}
\begin{eqnarray}
\frac{d \gamma}{d l} & = & 0    \nonumber   \\
\frac{d \lambda}{d l} & = & \frac{1}{2} \lambda   
\end{eqnarray}
    
   The above R. G. equations are equivalent to
\begin{eqnarray}
\frac{d \gamma}{d l}  & =  & - \gamma    \nonumber   \\
\frac{d \lambda}{d l} & =  & 0
\end{eqnarray}

   It is easy to see the fixed point is located at $ \gamma=0$
 where $ \widehat{a} $ loses its kinetic energy and becomes a Grassmann Lagrangian multiplier,
 it can be integrated out, therefore the impurity degree of freedoms
 completely disappear from the effective Hamiltonian and leave behind the 
 boundary conditions \cite{trick}
\begin{equation}
  b_{sf,L}(0)=-b_{sf,R}(0)
\label{bound2}
\end{equation}

   The boundary conditions Eqs.\ref{bound1},\ref{bound2} can be expressed
  in terms of bosons \cite{tough}
\begin{equation}
  \Phi_{s,L}(0)=\Phi_{s,R}(0)+\pi, ~~~ \Phi_{sf,L}(0)=-\Phi_{sf,R}(0)
\end{equation}

 The three Majorana fermions in the spin sector being twisted, the fixed point 
 possesses the symmetry $ O(3) \times O(5) $.
  Using totally different method, ML \cite{ludwig} identified the symmetry.
 By using the triality transformation, they
  showed that Eqs. \ref{bound1}, \ref{bound2}
  imply that the {\em original} fermions scatter
 into the collective excitations which fit into the spinor representation of
 $ SO(8) $.

 This symmetry enable us to work out
 easily the finite size spectrum at the fixed point in Table \ref{finite}.
 Although the conformal embedding is highly {\em non-trivial} in the original
 fermion basis \cite{affleck,twoimp}, it is completely {\em trivial} in the basis of
 Eqs.\ref{bound1},\ref{bound2},
 namely the finite size energy spectrum are simply constructed by the direct sum of 3 Majorana 
 fermions in R sector and  5 Majorana fermions in NS sector or {\em vice versa}.
 Table \ref{finite} is more compact than, but compatible with those listed
  in Ref. \cite{affleck,nrg}.

   By using Dyson equations, we can get all the correlation functions in $ (k, \omega) $
   space from the fixed point action where $\gamma=0 $ 
\begin{eqnarray}
\langle \widehat{a}(\omega) \widehat{a}(-\omega)\rangle & = & [\frac{t^{2}}{2} \int \frac{dk}{2\pi}
 \frac{1}{ i\omega - v_{F} k} ]^{-1}    \nonumber   \\
\langle b_{sf}(k_{1},\omega)b_{sf}(k_{2},-\omega) \rangle & = &
 \frac{2\pi \delta(k_{1}+k_{2})}{ i \omega-v_{F} k_{1}}    \nonumber   \\
  & + & ( \frac{t}{i \omega-v_{F} k_{1}}) (\frac{t}{ -i \omega-v_{F} k_{2}})
\langle \widehat{a}(\omega) \widehat{a}(-\omega)\rangle    \nonumber   \\
\langle b_{sf}(k,\omega)\widehat{a}(-\omega)\rangle & =  & \frac{t}{i\omega-v_{F} k} 
\langle \widehat{a}(\omega) \widehat{a}(-\omega)\rangle 
\end{eqnarray}   
   where $ t= i \frac{\lambda}{ 2 \sqrt{2 \pi a}} $.

%  The above equations are consistent with the expectation that the translation invariance
%along $x $ direction is broken
%due to the impurity at the boundary, but the translation invariance along $\tau $ direction
%remains.

   Paying special attention to the convergence factors when performing contour integrals,
   we get all the correlation functions in $ (x,\tau) $ space.
\begin{eqnarray}
  \langle b_{sf}( x_{1},\tau_{1} ) \widehat{a}(\tau_{2})\rangle & \sim &  \left \{ \begin{array}{ll}
  \langle b_{sf}( x_{1},\tau_{1} ) b_{sf}( 0,\tau_{2})\rangle_{0}   &  if  ~~~  x_{1} > 0    \\   
  - \langle b_{sf}( x_{1},\tau_{1} ) b_{sf}( 0,\tau_{2})\rangle_{0}   &  if ~~~  x_{1} < 0     
                       \end{array}    \right.     \nonumber   \\
  \langle b_{sf}( x_{1},\tau_{1} ) b_{sf}( x_{2}, \tau_{2})\rangle & = &  \left \{ \begin{array}{ll}
  \langle b_{sf}( x_{1},\tau_{1} ) b_{sf}( x_{2},\tau_{2})\rangle_{0}   & if ~~~ x_{1} x_{2} > 0    \\   
   - \langle b_{sf}( x_{1},\tau_{1} ) b_{sf}( x_{2},\tau_{2})\rangle_{0}   & if  ~~~  x_{1} x_{2} < 0     
                       \end{array}    \right.  
\end{eqnarray}
   where $ \langle\cdots\rangle_{0} $ means the non-interacting correlation functions.
   Any multi-point correlation functions can be calculated by Wick theorem.

    Integrating over $ k $, the correlation functions along the boundary follow \cite{trick}
\begin{eqnarray}
  \langle \widehat{a}(\tau_{1} ) \widehat{a}( \tau_{2})\rangle & = &  \frac{1}{\tau_{1}-\tau_{2}}   \nonumber   \\
  \langle b_{sf}( 0,\tau_{1} ) b_{sf}( 0, \tau_{2})\rangle & = &  0   \nonumber   \\
  \langle b_{sf}( 0,\tau_{1} ) \widehat{a}( \tau_{2})\rangle & = &  \frac{1}{t} \delta(\tau_{1}-\tau_{2})
\end{eqnarray}

 Because of the relations $ \psi(x)= \psi_{L}(x)$ if $ x > 0 $;
 $ \psi(x)= \psi_{R}(-x)$ if $ x < 0 $ \cite{trick},
 the above equations imply that at the fixed point, the impurity completely disappear and
 turns into one of the {\em non-interacting} primary fields of the fixed point
\begin{equation}
\widehat{a}(\tau) \sim b_{sf,0}(0,\tau)
\end{equation}

   The left moving and right moving parts are related by the boundary
 condition Eq. \ref{bound2} $ b_{sf,R}(-x,\tau) =-b_{sf,L}(x,\tau) $ if $ x< 0 $,
 any correlation functions can be mapped to the corresponding
 free fermion correlation functions. 

  From the boundary conditions Eqs \ref{bound1}, \ref{bound2}, it is easy to see that
  at the fixed point, L-L and R-R moving correlation functions are exactly the same with
  non-interacting case, however $ \langle\psi_{L} \psi_{R}^{\dagger} \rangle =0 $, namely the single
  particle $ S_{1} $ matrix vanishes, this " Unitarity Puzzle" was comprehensively
  explained by ML \cite{ludwig}.   

  Away from the fixed point, there is only one leading irrelevant operator 
 moving away from EK line $ -\lambda_{z}^{\prime} \widehat{a} \widehat{ b} a_{s}(0) b_{s}(0)
 \sim \cos\Phi_{sf}(0) \partial \Phi_{s}(0) $ \cite{line}, the first order correction
  to the single particle L-R Green function ( $ x_{1}>0, x_{2}<0 $ ) 
  due to this operator is  \cite{ludwig}
\begin{equation}
\langle \psi_{1 \uparrow}( x_{1},\tau_{1} ) \psi^{\dagger}_{1 \uparrow}( x_{2},\tau_{2} ) \rangle  
  \sim   \lambda^{\prime}_{z} ( z_{1}- \bar{z}_{2})^{-3/2} 
\end{equation}
     where $ z_{1}=\tau_{1}+i x_{1} $ is in the upper half plane,
     $ \bar{z}_{2}=\tau_{2}+i x_{2} $ is in the lower half plane. 

    By conformal transformation to finite temperature, the above Eq. turns to Eq.(3.23)
    of Ref.\cite{resistivity} where AL derived the low temperature $ \sqrt{T} $ behavior
    of the electron resistivity. Higher order corrections can be similarly performed, since
    they are just the evaluations of the multi-point correlation functions of the free chiral bosons;
    the low temperature expansion of the electron resistivity is
\begin{equation}
    \rho(T) \sim \frac{\rho_{u}}{2}(1+ T^{1/2} + T^{3/2} + T^{2} + \cdots )
\end{equation}

   There are two subleading operators with dimension 2 \cite{line},
\begin{eqnarray}
  :\widehat{a}(\tau) \frac{\partial\widehat{a}(\tau)}{\partial \tau}:  \sim
  :b_{sf,0}(\tau) \frac{\partial b_{sf,0}(\tau)}{\partial \tau}: = 
 : \cos2\Phi_{sf}(0,\tau):-\frac{1}{2}  
  : (\partial \Phi_{sf}(0,\tau))^{2} :     \nonumber  \\
  :a_{s}(\tau) \frac{\partial a_{s}(\tau)}{\partial \tau} +  
  b_{s}(\tau) \frac{\partial b_{s}(\tau)}{\partial \tau}:  =  
   :(\partial \Phi_{s}(0))^{2}:     ~~~~~~~~~~
\end{eqnarray}

    The first one moves along the EK line, second order perturbation in this dimension 2 operator leads to 
 $ \langle b_{sf}(0,\tau) b_{sf}(0,0) \rangle  \sim \frac{\gamma^{2}}{\tau^{3}} $,
  for large $ \tau $ \cite{exp}. The second one moves away from EK line. 
 It is easy to see that  
  both operators do not contribute to the one particle Green function. Straightforward extensions
  of CFT analysis in Ref. \cite{line} leads to the same conclusion.
  The contributions of these two operators
  to thermodynamic quantities have been discussed in detail in Ref.\cite{line}.

    The boundary OPE of the spin and flavor singlet pairing operator is
\begin{equation}
\psi_{1\uparrow}(z_{1}) \psi_{2 \downarrow} (\bar{z}_{2} ) =i (z_{1}-\bar{z}_{2})^{-1/2}
               :e^{-i \Phi_{c}(0,\tau)} : + \cdots 
\end{equation}

   The correlation function of the pairing operator $ {\cal O}= : e^{-i \Phi_{c}(0) } : $ is
\begin{equation}
   \langle {\cal O}^{\dagger}(\tau) {\cal O}(0) \rangle = \frac{1}{ \tau}
\end{equation}

     It implies a $ ln T $ divergent pairing susceptibility at the impurity site.

 In this paper, we discuss the 
   2CK fixed point with the symmetry
  $ O(3) \times O(5) $. The fixed point
 of two impurity, one channel Kondo model (2IK) possesses $ O(7) \times O(1) $
 \cite{ludwig,twoimp,gan}. Gan mapped the fixed point
 Hamiltonian of the 2IK to that of the 2CK \cite{gan}. 
 However, {\em different} canonical transformations
 are employed in the two models. In the 2IK, the impurity spin $ S^{z}_{+}= S^{z}_{1}+S^{z}_{2} $
 takes integer values $ -1, 0, 1$, therefore the canonical transformation
  $ U=e^{i S^{z}_{+} \Phi_{s}(0)} $ employed in Ref. \cite{gan}
 does {\em not} change the boundary conditions of  $ a_{s}, b_{s}$, this
 explains why the fixed point symmetry of the 2IK is $ O(7) \times O(1) $
 instead of $ O(3) \times O(5) $.  The two channel spin-flavor Kondo model discussed
  in Ref.\cite{sf} has the symmetry $ O(2) \times O(6) $. The two channel flavor
  anisotropic Kondo model always flows to the fixed point
  with the symmetry $ O(4) \times O(4) $ \cite{flavor}.
  Most recently, in reanalyzing the problem
  of an non-magnetic impurity hopping between two sites in a metal, the author
  found a {\em line of non-fermi liquid fixed points} with the symmetry
  $ U(1) \times O(1) \times O(5) $ which continuously 
  interpolates between the 2CK and the 2IK \cite{hopping} and an additional
  NFL fixed point which is the same as 2IK.
  The common feature of the models mentioned above is that the boundary
  interaction terms can be expressed in terms of Majorana fermions.
    This series has classified some of boundary fixed points
   corresponding to four pieces of bulk fermions.
 In general, twisting even (odd ) no. of Majorana fermions, we get
  fermi (non-fermi) liquid fixed points. From Table \ref{finite}, it is also easy
 to conclude that the finite size spectrum of twisting even (odd) no. of Majorana fermions
 is (not) equally spacing, therefore can (not) be fitted by a phase shift.
 Ref.  \cite{kim} studied a model
   where the one channel conduction electrons transforming
   as spin 3/2 representation of $ SU(2) $(therefore also 4 pieces of bulk fermions)
   couple to the spin $ s=1/2 $ impurity,
  the fixed point symmetry of this model remains an open question, but it
  certainly {\em  does not} fall into the above series, because the boundary
  interaction term of this model {\em cannot}
  be expressed in terms of Majorana fermions. 

  I thank D. S. Fisher, B. I. Halperin, A. L. Moustakas and D. Anselmi for helpful
  discussions. This research was supported by NSF Grants Nos. DMR 9106237 and DMR9400396.

\begin{table} 
\begin{tabular}{ |c|c|c|c| } 
 $ O(3) $ & $ O(5)$ & $ \frac{l}{v_{F} \pi}( E-\frac{3}{16}) $  & Degeneracy  \\  \hline
    R     &    NS      &        0                               &    2      \\  \hline
    NS    &    R       &       $ \frac{1}{8} $                    &    4      \\ \hline
    R     &    NS+1st  &       $ \frac{1}{2} $                    &    10      \\ \hline
  NS+1st  &    R       &       $ \frac{5}{8} $                     &    12      \\ \hline
  R+1st   &    NS      &           1                            &    6       \\
    R     &    NS+2nd  &           1                            &    20      \\   \hline
   NS     &    R+1st   &       $ 1+\frac{1}{8}$                   &    20      \\
   NS+2nd &     R      &       $ 1+\frac{1}{8}$                   &    12      \\  
\end{tabular}
\caption{ The finite size spectrum at the 2CK fixed point
 with the symmetry $ O(3) \times O(5) $. R is the Ramond ( periodic) sector, NS
 is the Neveu-Schwarz (anti-periodic) sector; R+1st( NS+1st ) is the first excited
 state in R (NS) sector, et. al. When counting the degeneracy of a energy level, we use
 the fact that
 the ground state degeneracy of $ N $ Majorana fermions
 in R sector is $ 2^{[N/2]} $. } 
\label{finite}
\end{table}


\begin{references}
\bibitem{atten} For simplicity, in the following, we don't distinguish between $ \psi_{s,i}(x) $
               and $\psi_{s} (x) $. The primes will also be omitted.
\bibitem{trick} The boundary conditions should be $b_{sf}(0,\tau)=0 $, however
   $ b_{sf}(0,\tau)=\frac{1}{2}(b_{sf,L}(0,\tau)+b_{sf,R}(0,\tau)) $, curious readers
   should check the derivation of mapping three dimensional Kondo problem to
    a one dimensional problem in Ref. \cite{affleck}.
\bibitem{tough} However, for more complicated cases, see Ref. \cite{hopping}.
\end{references}
\end{document}